\documentstyle{mn}
\newif\ifAMStwofonts
\ifoldfss
  \ifCUPmtlplainloaded \else
    \NewTextAlphabet{textbfit} {cmbxti10} {}
    \NewTextAlphabet{textbfss} {cmssbx10} {}
    \NewMathAlphabet{mathbfit} {cmbxti10} {} % for math mode
    \NewMathAlphabet{mathbfss} {cmssbx10} {} %  "   "    "
  \fi
  \ifAMStwofonts
    \ifCUPmtlplainloaded \else
      \NewSymbolFont{upmath} {eurm10}
      \NewSymbolFont{AMSa} {msam10}
      \NewMathSymbol{\upi}     {0}{upmath}{19}
      \NewMathSymbol{\umu}     {0}{upmath}{16}
      \NewMathSymbol{\upartial}{0}{upmath}{40}
      \NewMathSymbol{\leqslant}{3}{AMSa}{36}
      \NewMathSymbol{\geqslant}{3}{AMSa}{3E}

      \let\geq=\geqslant 
    \fi
  \fi
\fi % End of OFSS

\ifnfssone
  \newmathalphabet{\mathit}
  \addtoversion{normal}{\mathit}{cmr}{m}{it}
  \addtoversion{bold}{\mathit}{cmr}{bx}{it}
  \newmathalphabet{\mathbfit} % math mode version of \textbfit{..}
  \addtoversion{normal}{\mathbfit}{cmr}{bx}{it}
  \addtoversion{bold}{\mathbfit}{cmr}{bx}{it}
  \newmathalphabet{\mathbfss} % math mode version of \textbfss{..}
  \addtoversion{normal}{\mathbfss}{cmss}{bx}{n}
  \addtoversion{bold}{\mathbfss}{cmss}{bx}{n}
  \ifAMStwofonts
    \ifCUPmtlplainloaded \else
      \UseAMStwoboldmath
      \makeatletter
      \new@mathgroup\upmath@group
      \define@mathgroup\mv@normal\upmath@group{eur}{m}{n}
      \define@mathgroup\mv@bold\upmath@group{eur}{b}{n}
      \edef\UPM{\hexnumber\upmath@group}
      \new@mathgroup\amsa@group
      \define@mathgroup\mv@normal\amsa@group{msa}{m}{n}
      \define@mathgroup\mv@bold\amsa@group{msa}{m}{n}
      \edef\AMSa{\hexnumber\amsa@group}
      \makeatother
      \mathchardef\upi="0\UPM19
      \mathchardef\umu="0\UPM16
      \mathchardef\upartial="0\UPM40
      \mathchardef\leqslant="3\AMSa36
      \mathchardef\geqslant="3\AMSa3E

      \let\geq=\geqslant 
    \fi
  \fi
\fi % End of NFSS release 1

\ifnfsstwo
  \DeclareMathAlphabet{\mathbfit}{OT1}{cmr}{bx}{it}
  \SetMathAlphabet\mathbfit{bold}{OT1}{cmr}{bx}{it}
  \DeclareMathAlphabet{\mathbfss}{OT1}{cmss}{bx}{n}
  \SetMathAlphabet\mathbfss{bold}{OT1}{cmss}{bx}{n}
  \ifAMStwofonts
    \ifCUPmtlplainloaded \else
      \DeclareSymbolFont{UPM}{U}{eur}{m}{n}
      \SetSymbolFont{UPM}{bold}{U}{eur}{b}{n}
      \DeclareSymbolFont{AMSa}{U}{msa}{m}{n}
      \DeclareMathSymbol{\upi}{0}{UPM}{"19}
      \DeclareMathSymbol{\umu}{0}{UPM}{"16}
      \DeclareMathSymbol{\upartial}{0}{UPM}{"40}
      \DeclareMathSymbol{\leqslant}{3}{AMSa}{"36}
      \DeclareMathSymbol{\geqslant}{3}{AMSa}{"3E}

      \let\geq=\geqslant 
    \fi
  \fi
\fi % End of NFSS release 2

\ifCUPmtlplainloaded \else
  \ifAMStwofonts \else % If no AMS fonts
    \def\upi{\pi}
    \def\umu{\mu}
    \def\upartial{\partial}
  \fi
\fi

\title[RXTE observations of the Galactic microquasar XTE J1748-288]
{RXTE observations of Galactic microquasar XTE J1748--288 
during its 1998 outburst}
\author[M.Revnivtsev, S.Trudolyubov and K.Borozdin]
  {M. G.~Revnivtsev$^{1,2}$,
  S. P.~Trudolyubov $^{1,2}$ and
  K. N.~Borozdin $^{3,1}$
\\
  $^1$ Space Research Institute, RAS, Profsoyuznaya 84/32, 117810 Moscow, 
	Russia \\
  $^2$ Max-Planck-Institut f\"ur Astrophysik,
              Karl-Schwarzschild-Str. 1, 85740 Garching bei Munchen,
              Germany \\
  $^3$ Los Alamos National Laboratory, Los Alamos,
              87545 New Mexico, USA
      }

\date{Accepted       Received       in original form ??}

\pagerange{\pageref{firstpage}--\pageref{lastpage}}
\pubyear{1999}

\input epsf

\begin{document}

\maketitle

\label{firstpage}

\begin{abstract}

\large
We present an analysis of the {\it RXTE} observations of 
the recently discovered Galactic microquasar XTE J1748--288 
during its 1998 outburst.  General spectral and temporal 
properties of the source and their evolution were very typical 
for the Galactic black hole candidates (BHC) and, in particular, 
black hole X-ray Novae.

The spectral evolution of the source during the outburst can be considered
a sequence of qualitatively distinct states. During the
first observations, corresponding to the maximum of X-ray flux,
the spectrum of the source consisted of a dominating
hard power law component and a soft thermal component, which can be described
by the model of multicolor disk emission. The hard component contributed
$\geq$80\% to the X-ray luminosity in the 3-25 keV energy band.
Overall two-component spectral shape is an attribute of {\it very high}
state (VHS) observed previously in BHC, but the domination of hard component is
unusual.  Later on, as the X-ray source faded, its energy spectrum qualitatively
changed, showing {\it high} (HS) and then {\it low} (LS) states, both 
typical for black hole binaries. 

As the energy spectrum changed, the fast variability also evolved
dramatically. 
Initially the power density spectrum was formed by a dominating
band-limited noise component, QPO features at 20-30 Hz 
and at $\sim$0.5 Hz, and a very low frequency noise component.  
After a significant decrease of the contribution of the hard spectral
component the amplitude of 
the fractional variability decreased by an order of magnitude and
the PDS spectrum adopted a power-law shape with a broad QPO peak
around 0.03 Hz. When the system switched to the LS, the PDS shape
changed again and the QPOs have not been detected since.

When the source was observed in VHS, 
a clear correlation between QPO parameters and X-ray flux was seen.
Such a correlation gives an insight into our understanding of the accretion
process in X-ray black hole binaries. 
  
\end{abstract}


\large
\section{Introduction}

The XTE J1748--288 was discovered as a new transient source on June 4,
1998 by All Sky Monitor (ASM) aboard the Rossi X-ray Timing Explorer
{\it (RXTE)} observatory (Smith, Levine \& Wood 1998). The X-ray source 
was localized with an accuracy of 1 arcmin in multiple scans of the 
region by the PCA/{\it RXTE} experiment (Strohmayer \& Marshall 1998).
Observations in the radio band by VLA revealed the presence of an
unresolved radio source with a position consistent with one found by PCA:
R.A.=17$^h$48$^m$05$^s$.06, Dec=--28$^\circ$28$\arcmin$25$\arcsec$.8
(equinox 2000.0, position uncertainty 0.$\arcsec$6, Hjellming,
Rupen \& Mioduszewski 1998). Significant variability of 
the new radio source (Hjellming et al. 1998a; Fender \& Stappers 1998)
strongly supports its association with the X-ray transient.
In June 14.31 radio source became extended, with a proper motion of
20-40 mas/day (Rupen, Hjellming \& Mioduszewski 1998). The intrinsic 
velocity of the moving jet was higher than $0.93$c for the distance
$\geq$8 kpc derived from a 21 cm HI absorption measurement 
(Hjellming et al. 1998b). 
 
Two quasi periodical oscillations (hereafter QPO) with the central
frequencies $\sim$0.5 Hz and $\sim$32 Hz were found 
in the power density spectrum (PDS) of the source in
the PCA observation of June 6, 1998 (Fox \& Lewin 1998). 

The flux from the source was above ASM/{\it RXTE} detection limit 
until August of 1998.  Pointed instruments of RXTE - PCA and HEXTE - 
observed the source quasi evenly from June till September, obtaining
a good coverage of the whole outburst. Here we present the results of 
a spectral and timing analysis of these RXTE observations.
 
\section{Observations and analysis}
 
In our analysis we used all the publicly available data obtained from RXTE TOO
(Target Of Opportunity) archive including 21 pointed observations and
``slew'' parts of two PI restricted observations. The 23 observations quasi 
evenly cover the 1998 outburst of the source with a total exposure of
$\sim 80$ ksec. The list of observations is presented 
in Table \ref{obslog}. 
 
\begin{table}
\small
\caption{{\it RXTE} observations of XTE J1748--288 in 1998
\label{obslog}}
\begin{tabular}{crccc}
\hline
\#&Obs.ID&Date, UT&Time start&PCA Exp.\\
&& & &sec\\
\hline
1&30188-05-01-00$^a$&04/06/98&20:05:04&16\\
2&30188-05-02-00$^a$&05/06/98&03:03:44&64\\
3&30171-02-01-00&06/06/98&09:41:20&2655\\
4&30185-01-01-00&07/06/98&07:56:32&2944\\
      5&..-02-00&08/06/98&06:23:28&3027\\
      6&..-03-00&09/06/98&12:48:00&3729\\
      7&..-04-00&10/06/98&03:38:24&7721\\
      8&..-05-00&11/06/98&12:52:16&3439\\ 
      9&..-06-00&13/06/98&12:51:28&3114\\
     10&..-07-00&15/06/98&04:53:36&1795\\
     11&..-08-00&18/06/98&20:55:12&2327\\
     12&..-09-00&22/06/98&22:30:08&3210\\
     13&..-10-00&27/06/98&11:39:28&1647\\
     14&..-11-00&08/07/98&16:21:52&1295\\
     15&..-12-00&13/07/98&06:44:00&2056\\
     16&..-13-00&18/07/98&04:00:32&10585\\
     17&..-14-00&30/07/98&09:44:48&6841\\
     18&..-15-00&05/08/98&18:25:20&4333\\
     19&..-16-00&13/08/98&10:17:36&1565\\
     20&..-17-00&20/08/98&16:41:20&1704\\ 
     21&..-18-00&25/08/98&03:32:00&1785\\
     22&..-19-00&14/09/98&08:17:36&886\\
     23&..-20-00&26/09/98&03:29:04&10287\\
\hline
\end{tabular}
$^a$ -- Only the ``slew'' parts of these observations were public
at the time of our analysis
\end{table}
For data reduction we used the standard FTOOLS package. 
For the spectral analysis we used PCA data collected in the 3--25 keV
energy range. The response matrix was constructed for every single
observation using PCARMF v3.5. For the PCA background estimations 
we applied a Very Large Events (VLE)-based model. 
For some observations, where the standard VLE-based background subtraction 
was not good enough for our purposes, we used the activation component 
for background models separately 
(see details in the RXTE Cook Book, ``Using the Latest PCABACKEST'').
All spectra were dead time corrected.  
The analysis of the spectra of the Crab nebula confirmed that 
the uncertainties attributed to the response 
matrix do not exceed 0.5--1.0\%. To account for the uncertainty 
in the knowledge of the spectral response, a 1 per cent systematic error 
was added to the statistical error in each PCA channel.
 
For our timing analysis we used the PCA timing modes $Binned$, $Single
Binned$, $Event$ and $Good$ $Xenon$. Background subtracted light curves 
from $Standard 2$ mode data were used to generate the power 
density spectra for very low frequencies (below $10^{-2}$ Hz).

\begin{figure}
\epsfxsize 8cm
\epsffile[45 190 563 709]{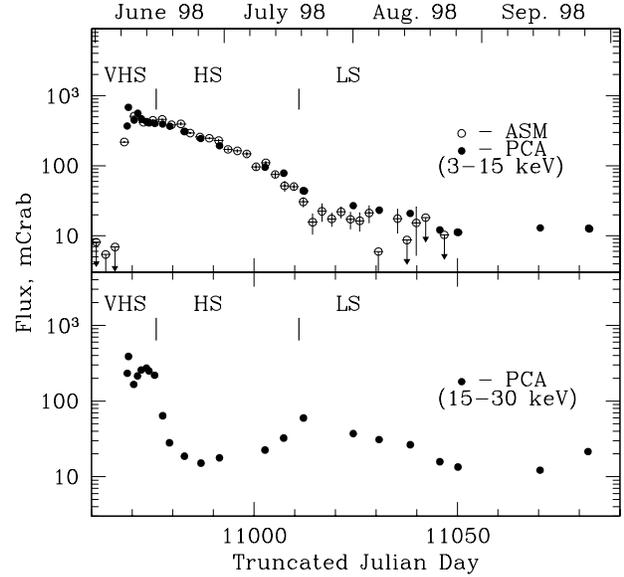}
\caption{The light curves of XTE J1748--288 during its 1998 outburst 
according to the data of RXTE/ASM ($2 - 12$ keV energy range -- upper panel, 
hollow circles) and PCA ($3 - 15$ keV energy range -- upper panel, filled 
circles; $15 - 30$ keV energy range -- lower panel, filled circles).
\label{lcurve} 
}
\end{figure}
 
Unfortunately, an analysis of the HEXTE data was strongly complicated by
the difficulties in the background subtraction. The source/background 
beam-switching of the HEXTE clusters 
(they ``rock'' periodically to $\pm$3.0$^\circ$ from the source
position), is not very efficient for precise background subtraction
when the source is located in the densely populated Galactic Center region.
There are a number of X-ray
sources around XTE J1748--288, namely GX 5-1, GRS 1758--258, 
GX 3+1 and 1E1740.7--2942 to mention few, each of them luminous enough 
to affect the results of the background measurement.
We have performed a check of the quality of the background
subtraction in any given observation by comparing the m-background
spectra (the spectra obtained during the -3.0$^\circ$ offset from the
source) with the p-background (+3.0$^\circ$ offset). Only observations 
with an adequate quality of subtraction were accepted for the analysis.
Because the rocking planes for clusters A and B are perpendicular 
to each other, different sky regions are within the field of view 
of these clusters during the background measurements.  
Therefore A clusters can provide more accurate background measurements 
for some observations, and B clusters - for others. 
After the 8-th observation,
when the source flux in the 20--100 keV energy band had dropped below $\sim$10
cnts/s/cluster, the accuracy of our knowledge of the background
became comparable to the source flux, and we excluded 
these later HEXTE data from our spectral analysis. 
However, we present the HEXTE observations of the source in its LS
in Fig.\ref{spectra}.  
The spectrum plotted there was obtained from carefully selected 
data, but still should be treated with care.
 
\section{Results}

\subsection{Energy spectra approximation}
We generated the energy spectra of XTE J1748-288 averaging the data 
over the whole single observations. To trace the evolution of the spectral
parameters within some observations, an additional analysis of the
spectra, accumulated in 256 sec time intervals, was performed. These
spectra were then used to study the correlation  of
the spectral and variability parameters (see section 3.4). 

For the spectral approximation of the first 15 observations
we used the model consisted of a multicolor disk blackbody emission 
component (Shakura \& Sunyaev 1973; Mitsuda et al. 1984), a power law 
component and low energy absorption.

Overall spectral shape for the remaining observations (16--23) can be
approximated by power law with low energy absorption. However, strong
emission line-like feature around $\sim$6.5 keV cannot be ignored. 
This feature may be approximated reasonably well by a Gaussian with 
centroid energy $E\sim6.5$ keV and width $\sigma\sim0.3$ keV.  
Its absolute intensity appeared to be stable against a
decrease of X-ray continuum by a factor of $\sim$4, thus we believe that
this line emission originates from the diffuse X-ray source near
the Center of the Galaxy rather than from XTE J1748-288 itself.
The measured intensity of this feature, $\sim3\times10^{-3}$
phot/s/cm$^2$, is in a reasonable agreement 
with previous measurements of the diffuse line emission from the Galactic
Center (e.g. Yamauchi \& Koyama 1993). The inclusion of the most
prominent Galactic diffuse lines -- 6.4, 6.7, and 6.9 keV -- also improves
the spectral approximation significantly (the widths of the lines were
frozen at 0.1 keV -- value comparable with the PCA energy resolution).  
The best fit line intensities ($\sim 1.3\times10^{-3}$ phot/s/cm$^2$ for 6.4
keV line and $\sim 1.5\times10^{-3}$ phot/s/cm$^2$ for 6.7 keV and 6.9
keV lines together \footnote{We added the fluxes from these 
two lines because PCA does not allow us to resolve them reliably})
were approximately constant within the accuracy of the measurements 
for all LS observations. The equivalent width of the whole
$K_\alpha$ feature increased from $\sim$300 eV in observation \#16
to $\sim$1.1 keV in observation \#23.

Spectral fits revealed the presence of an additional line in the spectrum,
with the central energy
around 8 keV and an intensity $\sim 2-3\times10^{-4}$ phot/s/cm$^2$.
Even given the moderate
energy resolution of the PCA, the presence of this line is evident
both when the Fe $K_\alpha$ complex is fitted by by three narrow lines 
as described above, and when one fits it by one broader line
with central energy $\sim$6.5 keV instead. The 8 keV line feature is
likely a complex of the Fe $K_\beta$ or/and Ni lines of the hard
($kT\sim 7$ keV) 
optically thin diffuse plasma emission of the Galactic Center region
\cite{gc_spectrum}. Note, that the similar spectral feature was
observed in the spectrum of XTE J0421+560/CI Cam \cite{Revnivtsev99},
that was likely caused by optically thin plasma thermal emission.

The results of the spectral approximation of the PCA data with the
analytical models described 
above are presented in Table 2. Fig.\ref{evol} shows an evolution of
some best fit parameters with time. A correlation between
changes in the soft fraction, the disk temperature and the power-law
slope during the first 15 observations (VHS) is clearly seen.

\subsection{Power density spectrum}

In order to analyze the timing properties of XTE J1748--288, 
we generated power density spectra (PDS) in the 0.01--250 Hz frequency 
range (for 2--13 keV energy band) using short stretches of data. The 
resulting spectra were logarithmically rebinned when necessary to 
reduce the scatter at high frequencies and normalized to the square of the
fractional variability. The white-noise level due to the Poissonian 
statistics corrected for the dead-time effects was subtracted (Vikhlinin, 
Gilfanov \& Churazov 1994; Zhang et al. 1995). To obtain PDS in the lower 
frequency band ($\sim 5\times10^{-4}$ Hz -- $\sim$0.01 Hz) we used the 16-sec
integrated data of {\em Standard 2} mode, because it allowed us to take into
account the influence of the background variation on the PDS, which
might be of importance at these frequencies. 

To study the evolution of the basic timing and spectral parameters of XTE
J1748--288 within the individual observations with a relatively high level of 
variability we generated power density spectra of the source averaged over
256 s time intervals according to the procedure described above (for
the results of this study see section 3.4).

We fitted the power density spectra of the source in the 0.02--150 Hz 
range to analytic models using the $\chi^{2}$ minimization technique to 
quantify their characteristics. For the approximation of the PDS obtained 
during the first 8 observations, 
we used the sum of a flat-topped band-limited noise (BLN) 
component: $P_{BLN}(f) \sim exp(-(f/f_{br})^{2}))$, where $f_{br}$ -- is a 
characteristic break frequency of the BLN continuum, a power law (PL)
component -- very low frequency noise (VLFN) and up to three harmonically
related Lorentzian groups to describe the QPO features (see explanation 
below).

\begin{figure}
\epsfxsize 8cm
\epsffile[45 190 563 709]{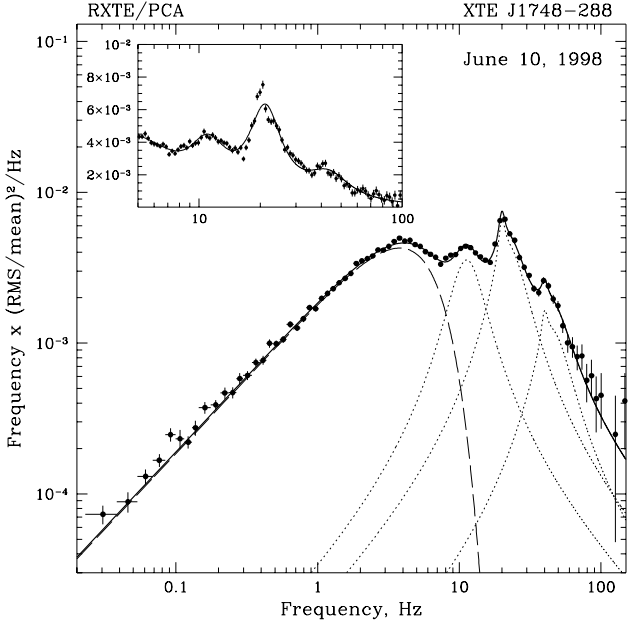}
\caption{Schematic presentation of the model used for the approximation of
the broad-band power density spectra of XTE J1748--288 during the period 
of maximum luminosity in units f$\times$(rms/mean)$^{2}$/Hz (thick solid 
line). The contributions of the band-limited noise components and the 
groups of Lorentzian components representing QPO features are shown by 
long-dashed and dotted lines respectively. The data shown are for 
the June 10, 1998 observation (PCA data, 2--13 keV energy range). 
The inset panel demonstrates the difference between the profile of the QPO 
peak and fit by single Lorentzian component. Inset axis units are the same 
as for the main panel. 
\label{pds_models} 
}
\end{figure}

Fig.\ref{pds_models} presents the data and model fit for one observation
(June 10). As is clearly seen from the inset of this figure, 
an approximation of the 20--30 Hz QPO features by a single Lorentzian
component does 
not provide an acceptable fit. This fact could be explained in terms 
of the existence of an additional high frequency shoulder, similar to 
those observed in Nova Muscae 1991 and GX 339-4 (Belloni et al. 1997). 
As we demonstrate below (section 3.4), the QPO centroid frequency derived from 
the simple Lorentzian fit shifts during the individual observations 
(Fig.\ref{corr1}). Since the strength of the QPO peak is anti-correlated 
with the QPO centroid frequency, averaging over the whole observation 
should lead to a steepening of the resulting QPO profile in its low 
frequency wing and a flattening in its high frequency wing forming a 
shoulder-like structure. Thus, the appearance of the observed QPO shoulder 
could be a result of such a displacement of the QPO centroid
frequency, for this reason it seems 
natural to approximate the QPO feature by a sum of several Lorentzians. 
To obtain a satisfactory fit we used the sum of two Lorentzians for every 
harmonic. The approximation shows that the frequencies of these additional 
Lorentzians are also harmonically related which supports our interpretation 
of the origin of the shoulders.

{\it In observations $\# 9 - 15$} the power density spectra 
were approximated by a power law component, the very low frequency noise 
(VLFN) component and a Lorentzian profile representing the low 
frequency QPO.

The best fit parameters of the band limited noise and QPO components are
presented in Table \ref{pds_params}. QPO rms amplitude is calculated 
as the quadratic sum of the corresponding rms amplitudes
for the Lorentzians used for the QPO approximation. 
Parameter errors correspond to the $1\sigma$ confidence level. These models 
approximate the data reasonably well, as indicated by the values of 
reduced $\chi^{2}$ for the fits.

\subsection{Evolution of the source during the outburst}

The X-ray flux histories of the XTE J1748-288 outburst based on the $RXTE$/PCA 
and ASM data in the 3--15 and 15--30 keV energy ranges are shown 
in Fig.\ref{lcurve}. The evolution of the source flux in the soft 
X-ray band (3--15 keV) was characterized by the fast initial rise to 
a level of $\sim$600--700 mCrab on a time scale of $\sim$2--3 days 
followed by the $\sim 8$ day--long maximum and a nearly exponential decay 
to the quiescent level with e-folding time $\sim 15$ days. The flux in
the hard band (15--30 keV) undergone a fast initial 
rise to the $\sim$300--400 mCrab level followed by the maximum phase 
of a $\sim8$--day duration and an abrupt decay, which is similar to the 
behavior of Nova Muscae GS/GRS 1124-683 
\cite{Ebisawa94} during its outburst of 1991.
Based on the results of the spectral and timing analysis, the evolution 
of XTE J1748--288, in analogy to Nova Muscae 1991, could be
divided into three distinct parts, corresponding to three main phenomenological
states: {\it very high} (VHS), {\it high} (HS) and {\it low} (LS).  
At the same time we would like to note, that in the first
8 observations the spectrum of XTE J1748--288 have an extremely bright 
hard component (the power law component contribution to 
the total 3--25 keV flux $\ga$80\%) which is somewhat different 
from the usual VHS spectrum. 
Very similar spectra were observed only with Ginga
at the very beginning of the outburst of GS 1124--683 
(Kitamoto et al. 1992; Ebisawa et al. 1994), and with Mir-Kvant 
at the beginning of the outbursts of X-ray Novae KS1730--312 
\cite{bor95,Trudolyubov96} and GRS 1739--278 \cite{bor98}, but
observations of XTE J1748--288 outburst provided 
an unique opportunity to study this unusually hard VHS spectrum 
in detail.  It is also worth mentioning that
contrary to many previous observations of the VHS, which
could be distinguished from the HS according to the difference in
the power spectrum only, in our case both the energy spectrum and
power spectrum in VHS were distinctly different from those in the HS.

\begin{figure}
\epsfxsize 8cm
\epsffile[45 190 563 709]{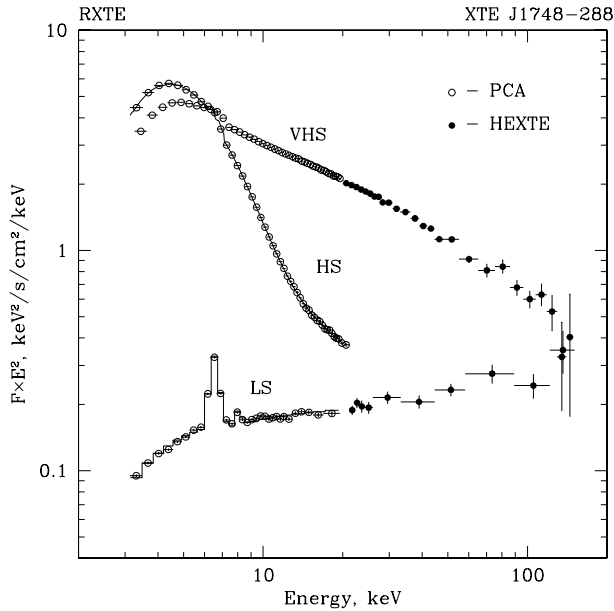}
\caption{Typical broad-band energy spectra of 
XTE J1748--288 during the different stages of the 1998 
outburst: $VHS$ -- observation \#4, $HS$ -- observation \#10
 and $LS$ -- (observation \#20). Hollow and filled circles
represent the data of PCA and HEXTE instruments respectively.
\label{spectra} 
}
\end{figure}

\begin{figure}
\epsfxsize 8cm
\epsffile[45 190 563 670]{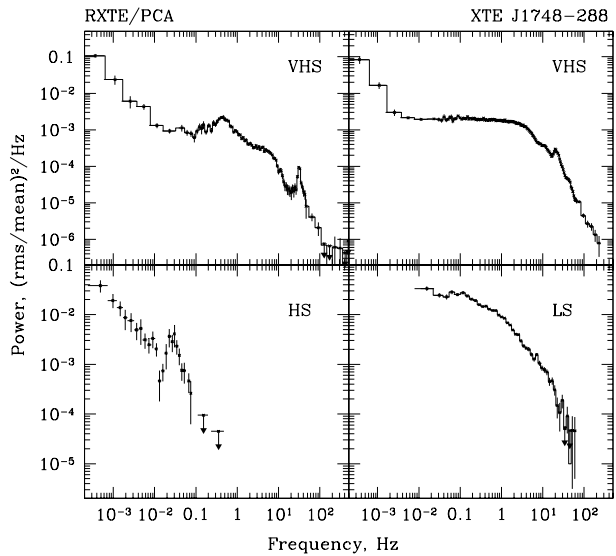}
\caption{Typical broad-band power density spectra of XTE J1748--288 
for different spectral states of the source: {\em VHS} --
averaged observations \#3--4 (upper lest panel), {\em VHS}
-- averaged observations \#5--8 (upper right panel), {\em HS}
-- averaged observations \#9--15 (left lower panel) and  {\em LS} 
-- averaged observations \#16--20 (right lower panel).   
PCA data for 2 -- 13 keV energy band. 
\label{pdss} 
}
\end{figure}

\begin{figure}
\epsfxsize 8cm
\epsffile[45 190 563 700]{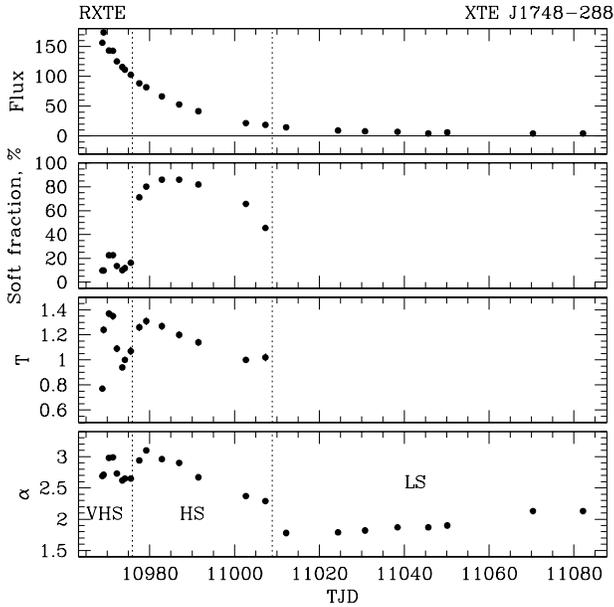}
\caption{The evolution of the spectral parameters of 
multicolor blackbody disk plus the power law model
for the XTE J1748--288 observations. The soft
fraction denotes the ratio of the flux of the soft component to the total
flux from the source in energy band 3--25 keV. The dotted lines 
on the panels denote the approximate time of spectral transitions. 
\label{evol} 
}
\end{figure}

\begin{enumerate}
 
\item[{\em Very High State}] During the first eight RXTE 
pointed observations (June 4 -- 11, 1998) 
the source was found in an extremely bright state. The broad-band 3--150 
keV energy spectrum of XTE J1748--288 in this state can be satisfactorily 
described by the sum of two components: a relatively weak soft thermal 
component (contributing only $\sim$ 10--20 per cent to the total X-ray flux 
in the 3--25 keV energy band) with a characteristic temperature of
$\sim$ 0.8--1.4 keV and a strong hard component, which is approximately 
a power law shape and does not show 
the evidence of a high energy cut-off up to $\sim 150$ keV \footnote{
An averaged broad-band 3--150 keV spectrum of the observations 
\#3--8 is suggestive for the presence of an additional 
spectral component in the 10--40 keV energy range, 
which can be fitted by the Compton reflection (from the neutral
medium) model with the reflection parameter $\Omega / 2 \pi \sim 0.2$. 
The presence of this component is not required if one analyses 
the PCA and HEXTE data separately. The cross-calibration
uncertainties between the two instruments do not allow us to state conclusively
regarding the presence of this component in the spectrum}. 
The corresponding luminosity in the 3--25 keV band uncorrected
for the interstellar absorption 
was at a level of 1--1.5$\times10^{38}$ ergs/s (assuming a 8 kpc 
distance) during this period.

The broad-band power density spectrum (PDS) of the source is formed by 
the dominating band-limited noise component (corresponding rms amplitude 
$\sim$ 5--9 \%), the strong QPO feature with centroid frequency 
$\sim$ 20--30 Hz and a very low frequency noise (VLFN) component with 
a slope $\sim - 1.5$ and rms amplitude $\sim$ 1.0-1.5 \% 
(Fig.\ref{pdss}, {\it upper panels}).

The correlated spectral and timing evolution of the source during this
period is of particular interest. The local increase of the contribution of the 
soft spectral component to the total X-ray flux in observations 
\# 3--4 (June 6 -- 7, 1998) was accompanied by a steepening of the hard 
spectral component and significant changes in the power density spectrum 
(Table 2, 3. Fig.\ref{pdss}, left upper panel), i.e.:
\par 
-- the increase of the BLN characteristic break frequency 
and the decrease of its rms amplitude; 
\par 
-- the increase of the QPO centroid frequency from $\sim 20$ Hz to 
$\sim 30$ Hz; 
\par 
-- the appearance of the additional broad QPO feature at $\sim 0.5$ Hz 
(Fox \& Lewin 1998); 
\par 
-- the notable rise of the VLFN rms amplitude.
\par
Detected correlations between the spectral and timing properties 
of the source in this state, discussed in more detail below (see \S3.4), 
holds a wide range of the time scales from seconds to several days. 

The two-component spectrum, consisting of the soft thermal component and 
a hard power law component is a signature of the
{\it high} spectral state of the Galactic black hole candidates 
(see Tanaka \& Shibazaki 1996, and references therein).  The {\it very high}
state differs from {\it high} state mostly by the much stronger 
fast variability,
while the differences in energy spectra are not significant and 
no good criterium was suggested to distinguish between these two
states on spectral ground. In case of XTE J1748-288
the spectrum was sufficiently different from the typical spectra
observed previously in the HS and the VHS,
because power law component dominated in the 3--25 keV energy band.  Fast
variability and shape of power spectrum was typical for the VHS
of Galactic black hole binaries (e.g. Belloni et al. 1997).

\item[{\em High State}]. As the flux of XTE J1748--288 dropped 
below $\sim$4200 cnts/s/PCA (observations \#9--\#15), the spectral 
and timing properties of the source changed drastically. The strength 
of the hard component in the energy spectrum decreased rapidly and 
the contribution of the soft thermal component to the total luminosity 
in the 3--25 keV energy range increased to more than 60 per cent as 
typical for HS (Fig.\ref{spectra}, \ref{evol}; Table 2). 

Simultaneously the shape of the power density spectrum changed qualitatively.
The band-limited noise (BLN) component and the high frequency 20--30 Hz 
QPOs became undetectable, and the amplitude of the fractional variability 
decreased from $\sim$10 to $\sim$1 per cent. The power density spectrum during 
this period was composed of the power law component with a slope of --1.0--1.5 
and a broad QPO peak at $\sim$0.03 Hz (see Fig.\ref{spectra},\ref{evol}; 
Table 3). 

\item[{\em Low State}]. The subsequent decline in the intensity of the
XTE J1748--288 X-ray flux was accompanied by a gradual decrease 
of the strength of the soft spectral component (Table \ref{params}) 
and in 16th observation (July 18, 1998) the source was found already 
in the standard LS. It should be noted that in this state 
the source flux was extremely weak and a substantial part of the observed 
emission in the PCA energy band can be attributed to the galactic diffuse 
and point sources, in particular to Sgr B2 (the offset from XTE J1748--288 
$\sim$12$\arcmin$).

The {\it low} state X-ray spectrum in the 3--25 keV energy 
band was composed by a power law component with a
photon index of $\alpha\sim$2, a complex of strong emission lines 
(around 6.5 keV and at $\sim$8 keV) and low energy absorption.  
The absolute intensity of the lines
appeared to be very stable against the decrease of X-ray continuum, which
strongly supports the interpretation that they originate from some 
of the sources within the field of view, but not from the XTE J1748--288.
We have performed an additional study to get an estimation for 
the intensity of the emission line attributed to the XTE J1748--288 itself,
assuming the centroid energy of the intrinsic source line to be equal 
to 6.4 keV. 
Using the fact that the strong decline in the continuum flux did not 
result in the statistically significant change in the observed 6.4 keV
line flux, we obtained 2$\sigma$ upper limit $\sim$80 eV
on the equivalent width of the intrinsic emission line. 
 
\end{enumerate}
 
\begin{table*}
\small
\caption{Spectral parameters of XTE J1748--288, derived using the
combination of the multicolor disc blackbody (Mitsuda et al. 1984) 
and power law models with correction to the interstellar absorption. 
For the low state observations a power law approximation was used. 
Parameter errors correspond to $1 \sigma$ confidence level. 
\label{params}}  
\begin{tabular}{rcccccccc}
\hline
\#&$T_d$&$R_{eff}\sqrt{cos~i}~D_{10}$$^*$&$\alpha_{pl}$&$N_H$&Flux (3--25 keV)&Flux$_{\rm soft}$(3-25 keV)&$\chi^2$(46 dof)\\
  &keV  & km      &ph. index&$10^{22}$ $cm^{-2}$&$\times 10^{-10}$
ergs/s/cm$^2$&$\times 10^{-10}$ ergs/s/cm$^2$\\
\hline
\multicolumn{7}{c}{Very High State}\\
\hline
1 &$0.77\pm0.02$&$55.29\pm1.11$&$2.69\pm0.03$&$10.9\pm2.1$&$156.4\pm4.7$&$15.2\pm0.3$&29.54\\
2 &$1.24\pm0.03$&$11.95\pm0.89$&$2.71\pm0.03$&$8.17\pm0.8$&$173.8\pm5.3$&$16.9\pm0.5$&34.08\\
3 &$1.37\pm0.03$&$13.17\pm0.26$&$2.98\pm0.03$&$10.4\pm0.3$&$143.2\pm4.3$&$32.3\pm0.4$&29.24\\
4 &$1.35\pm0.03$&$13.47\pm0.27$&$2.99\pm0.03$&$10.1\pm0.3$&$142.8\pm4.3$&$32.4\pm1.0$&26.97\\
5 &$1.09\pm0.03$&$17.65\pm0.35$&$2.73\pm0.03$&$9.3\pm0.3$&$125.0\pm3.8$&$17.0\pm1.0$&28.83\\
6 &$0.94\pm0.02$&$23.27\pm0.47$&$2.62\pm0.03$&$9.3\pm0.4$&$115.6\pm3.5$&$11.7\pm0.5$&31.26\\
7 &$1.00\pm0.02$&$20.44\pm0.41$&$2.65\pm0.03$&$9.1\pm0.3$&$111.2\pm3.3$&$13.1\pm0.4$&25.95\\
8 &$1.07\pm0.03$&$17.93\pm0.36$&$2.65\pm0.03$&$7.3\pm0.3$&$102.6\pm3.1$&$16.7\pm0.4$&16.68\\
\hline
\multicolumn{7}{c}{High State}\\
\hline
9 &$1.26\pm0.03$&$21.27\pm0.43$&$2.94\pm0.03$&$6.8\pm0.2$&$88.4\pm2.7$&$62.9\pm0.5$&26.63\\
10 &$1.31\pm0.03$&$19.92\pm0.40$&$3.10\pm0.03$&$7.8\pm0.3$&$81.8\pm2.5$&$65.6\pm1.9$&30.54\\
11 &$1.27\pm0.03$&$20.39\pm0.41$&$2.96\pm0.03$&$7.5\pm0.2$&$66.5\pm2.0$&$57.2\pm2.0$&24.35\\
12 &$1.20\pm0.03$&$21.07\pm0.42$&$2.90\pm0.03$&$7.4\pm0.2$&$52.9\pm1.6$&$45.5\pm1.7$&31.80\\
13 &$1.14\pm0.03$&$21.08\pm0.42$&$2.67\pm0.03$&$7.4\pm0.2$&$41.5\pm1.2$&$34.0\pm1.4$&30.40\\
14 &$1.00\pm0.02$&$20.05\pm0.40$&$2.37\pm0.03$&$6.8\pm0.3$&$21.6\pm0.6$&$14.2\pm1.0$&34.94\\
15 &$1.02\pm0.03$&$13.95\pm0.28$&$2.29\pm0.03$&$5.7\pm0.3$&$18.7\pm0.6$&$8.5\pm0.4$&36.11\\
\hline
\multicolumn{7}{c}{Low State}\\
\hline
%\#& & &$\alpha$&$N_H$&PL flux (3-25 keV)& & $\chi^2$(46 dof)\\
%& & &ph. index&$10^{22}$ cm$^{-2}$&$\times 10^{-10}$ ergs/s/cm$^2$ & \\
%\hline
16& & &$1.78\pm0.02$&$3.4\pm0.4$&$14.6\pm0.7$ & &32.2\\
17& & &$1.79\pm0.02$&$4.9\pm0.5$&$9.2\pm0.6$  & &47.0\\
18& & &$1.82\pm0.02$&$5.0\pm0.2$&$7.8\pm0.5$  & &33.1\\
19& & &$1.87\pm0.02$&$5.8\pm0.3$&$7.0\pm0.4$  & &30.6\\
20& & &$1.87\pm0.02$&$5.8\pm0.2$&$4.2\pm0.3$  & &28.1\\
21& & &$1.90\pm0.02$&$5.2\pm0.5$&$6.0\pm0.3$  & &53.5\\
22& & &$2.13\pm0.02$&$7.3\pm0.4$&$4.2\pm0.2$  & &48.6\\
23& & &$2.13\pm0.02$&$7.0\pm0.5$&$4.1\pm0.1$  & &51.3\\
\hline
\end{tabular}
\par
$^*$--$D_{10}$ -- source distance in units of 10 kpc. $i$ -- inclination
angle of the system

\end{table*}

\begin{table*}

\caption{The characteristics of the power density spectra of XTE J1748--288.
Parameter errors correspond to $1 \sigma$ confidence level. rms$_{total}$, 
and rms$_{BLN}$ represent total rms amplitude and total rms amplitude of 
the band-limited noise (BLN) component integrated over $0.02 - 150$
Hz frequency range, f$^{br}_{BLN}$ and f$_{QPO}$ represent the
characteristic break frequency of the BLN component and the QPO centroid
frequency. rms$_{QPO}^{1}$, rms$_{QPO}^{1/2}$ and rms$_{QPO}^{2}$ are the
rms amplitudes of the groups of Loretzians used to approximate the
first (fundamental) harmonic, subharmonic and second harmonic of the
QPO. f$_{QPO}^{shoulder}$ represents the centroid frequency of the
accompanying Lorentzian (see text)
\label{pds_params}}
\begin{tabular}{cccccccccc}
\hline
$\#$&${\rm rms_{total}}$ &${\rm f^{br}_{BLN}}$, Hz&${\rm rms_{BLN}}$&${\rm f_{QPO}}$, Hz&${\rm f_{QPO}^{shoulder}}$&${\rm rms_{QPO}^{1}}$&${\rm rms_{QPO}^{1/2}}$& ${\rm rms_{QPO}^{2}}$&$\chi^{2}$(dof) \\
\hline
\multicolumn{10}{c}{Very High state (2--13 keV, 0.02--150 Hz)}\\
\hline
2&$11.7\pm0.3$&$3.79\pm0.32$&$8.69\pm0.71$&$21.8\pm0.5$& - &$5.8\pm0.9$&$4.7\pm0.9$&-&$33(36)^*$\\
3&$7.4\pm0.1$&$6.14\pm0.14$&$5.15\pm0.16$&$31.6\pm0.2$& - &$3.2\pm0.1$&    -         &-&$228(183)$\\
 &&             &             &$0.48\pm0.02$& &$2.7\pm0.2$\\
4&$7.1\pm0.1$&$6.15\pm0.13$&$4.83\pm0.14$&$31.3\pm0.2$& - &$3.1\pm0.2$&    -         &-&$229(183)$\\
 &&             &             &$0.44\pm0.02$& &$2.9\pm0.2$\\
5&$12.1\pm0.1$&$4.27\pm0.05$&$8.80\pm0.11$&$23.7\pm0.1$&$24.8\pm0.2$&$5.4\pm0.3$&$5.0\pm0.2$&$2.5\pm0.4$&$217(177)$\\
6&$13.2\pm0.1$&$3.68\pm0.04$&$8.95\pm0.09$&$20.2\pm0.1$&$21.3\pm0.1$&$6.6\pm0.3$&$5.6\pm0.1$&$3.3\pm0.3$&$208(177)$\\
7&$13.1\pm0.1$&$3.79\pm0.03$&$9.04\pm0.05$&$20.0\pm0.1$&$21.5\pm0.1$&$6.3\pm0.3$&$5.5\pm0.1$&$3.3\pm0.3$&$209(177)$\\
8&$12.4\pm0.1$&$4.10\pm0.04$&$8.91\pm0.11$&$22.6\pm0.2$&$24.1\pm0.2$&$5.5\pm0.3$&$5.2\pm0.2$&$3.0\pm0.3$&$188(177)$\\
\hline
\multicolumn{10}{c}{High state (2-13 keV, $5\times 10^{-4}$--0.2 Hz)}\\
\hline
 && $\alpha_{\rm VLFN}$&${\rm rms_{VLFN}}$,\% & ${\rm f_{QPO}}$, Hz&FWHM, Hz&rms$_{\rm QPO}$,\%&&&$\chi^2$(dof)\\
\hline
9--15&&$-1.14\pm0.08$&$0.86\pm0.09$&$(2.9\pm0.1)\times10^{-2}$&$(1.1\pm0.4)\times10^{-2}$&$0.77\pm0.14$&&&21.1(24)\\
\hline
\end{tabular}

\medskip
-- all $rms$ values in \%

$^*$ -- this power density spectrum was obtained in the frequency range
0.3--50 Hz (slew part of the observation with total exposure 64 sec)
\end{table*}

\subsection{Correlations between spectral and variability parameters in the 
very high state}
\begin{figure}
\epsfxsize 8cm
\epsffile[45 190 563 709]{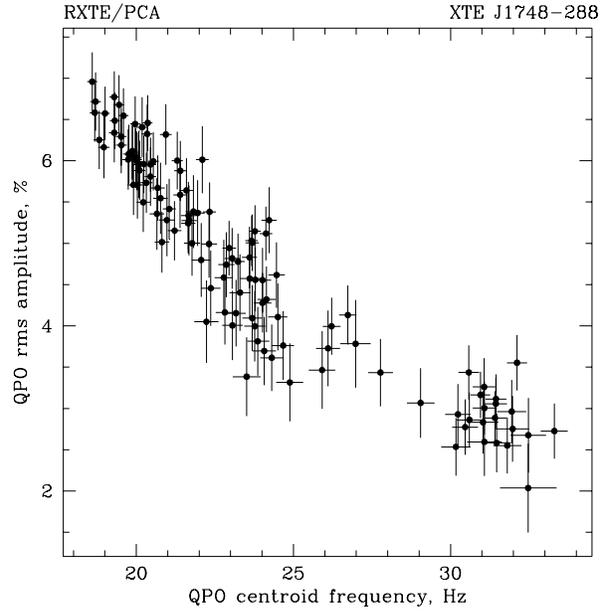}
\caption{The dependence of the rms amplitude of the high frequency QPO peak 
on its centroid frequency in the VHS observations 
(June 4 -- June 11, 1998; PCA data). Each point corresponds to the 
data averaged over 256 sec time intervals.
\label{corr1} 
}
\end{figure}

The significant evolution of the spectral and temporal properties of 
XTE J1748--288 in the VHS provides an excellent 
opportunity to study their mutual relationship in a wide range of 
time scales from seconds to several days. In the course of our extended 
timing analysis the general correlation between the main parameters 
describing the power density spectrum of the source (QPO centroid 
frequency, characteristic break frequency of the BLN component, 
BLN rms, QPO rms) was observed. The dependence of the 
rms amplitude of main QPO peak on its centroid frequency is shown in 
Fig.\ref{corr1} .  The plot demonstrates evident anticorrelation 
between these two parameters.  Other PDS parameters also appeared 
to be strongly inter-correlated.

\begin{figure}
\epsfxsize 8cm
\epsffile[45 190 563 709]{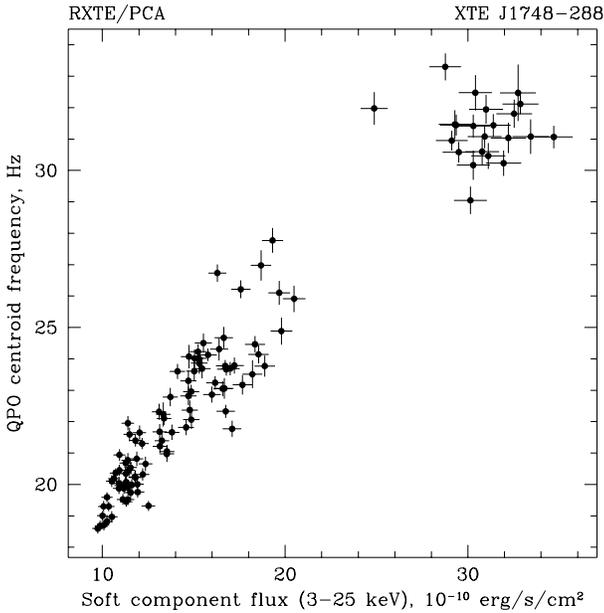}
\caption{The dependence of the QPO centroid frequency on the X-ray flux in the
soft spectral component in the 3--25 keV energy range
(without correction for an interstellar absorption).  The parameters of
the soft component are for the best-fit approximation by 
multicolor disk black body plus power law model. PCA data 
for the VHS observations (June 4--11, 1998)
have been used. Each point corresponds to the data averaged over
256-sec time intervals.
\label{corr} 
}
\end{figure}

Most striking is the established close relation between the 
evolution of the spectral and timing parameters of the source. 
The change of the QPO centroid frequency is correlated with change 
of the spectral parameters, derived from the energy spectra fits. 
In particular, there is a clear trend of increasing the QPO centroid 
frequency with rise of the soft component flux (Fig.\ref{corr}). 
It should be noted that this type of correlation holds on the wide 
range of time scales (from seconds to several days).

A similar type of the correlation between the QPO parameters and X-ray flux
has also been reported for other Galactic superluminal jet source 
GRS 1915+105 (Trudolyubov, Churazov \& Gilfanov 1999; Markwardt, 
Swank \& Taam 1999), X-ray Nova XTE J1550--564 (Cui et al. 1999), and 
XTE J1806--246, that is suspected to be a system containing a neutron star 
(Revnivtsev, Borozdin \& Emelyanov 1999). 
As in the cases of GRS 1915+105 and XTE J1550--564 the QPO peak 
in the power density spectrum of XTE J1748--288 was detected 
only in observations characterized by 
the domination of the hard spectral component. This fact hints on the
direct link between the processes of QPO and hard  
spectral component formation.

\section{Discussion} 
 
Many bright X-ray transients are reputed black hole candidates and
demonstrate similar X-ray spectra and fast variability. During the 
outburst these systems are typically found in one of two qualitatively 
distinguishable spectral states: in the {\it high} state composed of  
the bright 
thermal component and the extended hard power-law, or in the {\it low} state 
with the hard power-law spectrum with a photon index  of $\sim$1.5 and an 
exponential high energy cutoff. A more detailed description of these 
states can be found elsewhere \cite{sun94,TL95,Tanaka96}. 
A third, {\it very high}, state has been recognized also, with two 
component spectrum similar to the HS, but with somewhat stronger power-law
component and with much stronger fast variability 
(Miyamoto et al. 1991, 1994; Takizawa et al. 1997). The spectral 
evolution of X-ray transients is usually in correlation with X-ray flux 
changes, so it is widely believed that state transitions are driven 
by the variable accretion rate. Such transitions were also observed in 
the persistent black hole systems, namely, Cyg X-1 and GX 339-4, but the
dynamics of these systems is much slower, so they are more often 
observed in one of these states and typically switch to another state 
once every several years \cite{sun79,mak86,cui98,dove98,Trud98}.

The transient X-ray source XTE J1748--288 resembles other black hole 
X-ray transients in many ways. The general properties of the source 
light curves are similar to the properties of the light curves of black 
hole X-ray Novae such as A 0620-00, GS 2000+25, GS/GRS 1124-68, 
GRS 1009-45, GRS1739-278 and others (see Chen, Shrader \& Livio 1997 
for review, and references therein). During the 1998 outburst the source was 
observed by {\it RXTE} in {\it very high}, {\it high} and {\it low}
spectral states, which are very typical for black hole
systems. During several observations 
corresponding to the peak X-ray flux, the spectrum of the source was 
in an unusual VHS, a very bright hard component
was observed. Similar spectra were 
detected at the beginning of the outbursts of X-ray black hole Novae 
GS/GRS1124--68 \cite{Ebisawa94,Miyamoto94}, 
KS/GRS 1730--312 \cite{bor95,Trudolyubov96} and GRS 1739--278 
\cite{bor98}. Here we studied this spectrum in more detail.

The transition of XTE J1748--288 from the {\it very high} state to 
the {\it high} state changed both the spectral and
fast variability properties of the X-ray flux of XTE
J1748--288. Unlike the case of Nova Muscae
\cite{Miyamoto94,Ebisawa94}, hard component dominated spectrum 
during all VHS observations.
The soft component in the spectrum of the
XTE J1748--288 increased rapidly by a factor of $\sim$4 after the
transition, while the total flux in the energy band 3--25 keV
continued to decrease smoothly. The amplitude of the fast
variability dropped down during the transition and PDS shape
changed qualitatively (see Fig.\ref{pdss}). PDS of the source X-ray
flux in the HS has a power law shape of 
the continuum and shows the 
presence of the low-frequency QPO peak at $\sim 0.029$ Hz, resembling 
the properties of the {\it high} state PDS of the well-known 
black hole candidate LMC X-1 (its form was described by the sum 
of the power law component and a QPO peak with a centroid frequency of
$\sim 0.075$ Hz) (Ebisawa et al. 1989). 

After the subsequent decrease of the luminosity, the XTE J1748--288 
switched the 
source to the {\it low} state typical for the black hole X-ray 
Novae at this stage of the outburst evolution.

The shape of the broad-band power density spectrum of XTE J1748--288 in 
the VHS was similar to the PDS observed for the Galactic black hole 
candidates GX 339-4, Nova Muscae 1991 and  GRS 1915+105 (Miyamoto et
al. 1994; Belloni et al. 1997; Trudolyubov, Churazov \& Gilfanov 1999): 
it was dominated by a strong band-limited noise with QPO components. 

The correlated spectral and timing evolution of the source during the 
VHS and, in particular, the observed trend of increasing the QPO centroid 
frequency with the rise of the soft component flux on the wide range of 
time scales is of special importance.  A similar type of correlation 
between the QPO parameters and X-ray flux have also been reported for 
other Galactic black hole candidates: GRS 1915+105 and XTE J1550--564 
(Trudolyubov, Churazov \& Gilfanov 1999; Markwardt, Swank \& Taam 1999; 
Cui et al. 1999), which hints at the general mechanism of the X-ray 
emission generation for this class of objects.

One of the possible and probably the most promising way to explain 
these observational results for black hole systems 
is to use the models based on the idea of a shock front formation
between two distinct parts of the accretion flow producing 
the bulk of X-rays: the central region responsible for the generation of 
the hard spectral component and optically thick accretion disk
(Chakrabarti \& Titarchuk 1995). The oscillations in shock 
geometrical parameters (i.e. height, width) may modulate the flux 
of the ``seed'' photons from the inner boundary of the optically thick 
accretion disk and change the state of matter in the optically thin region 
- and as a result, QPOs of X-ray flux are to be detected
(Titarchuk, Lapidus \& Muslimov 1998; Molteni, Sponholz \& Chakrabarti, 1996). 
In this case the time scale of the oscillations is determined 
by the position of the boundary between the optically thick and thin regions, 
which in turn is directly related to the luminosities of the soft and 
hard spectral components. Hence the observed correlation between the QPO 
frequency and X-ray flux could be naturally explained in the framework of 
this model.

The observed complex of the general spectral and timing properties of 
XTE J1748--288 is similar to that of the dynamically proven black 
hole X-ray Novae and Galactic superluminal jet sources. This fact 
could hint at the universal character of evolution of the accretion 
flow in the black hole systems and its direct link to the mechanism 
of the formation of relativistic ejections of matter. 

\section*{Acknowledgments}
This research has made use of data obtained through the High Energy
Astrophysics Science Archive Research Center Online Service, provided 
by the NASA/Goddard Space Flight Center. 

The authors wish to thank Dr.M.Gilfanov for his valuable suggestions, 
which helped us to improve the paper significantly. We are grateful 
to M.N.R.A.S. anonymous referee for his/her helpful comments, and
to Ms. Kate O'Shea for the language editoring of the manuscript.
Part of this work was supported by RBRF 96-02-18588
and INTAS 93-3364-ext grants.

\label{lastpage}


\begin{thebibliography}{99}
\bibitem[Belloni et al. 1997]{bel97} Belloni T., van der Klis, M., 
Lewin W.\ H.\ G., van Paradijs J., Dotani T., Mitsuda K., Miyamoto S., 
1997, A\&A, 332, 857 

\bibitem[Borozdin et al. 1995]{bor95} Borozdin K.\ N., Aleksandrovich N.\ L., 
Arefiev V.\ A., Sunyaev R.\ A., Skinner G.\ K., 1995, Astr.L. 21, 212

\bibitem[Borozdin et al. 1998]{bor98} Borozdin K.\ N., Revnivtsev M.\ G.,
Trudolyubov S.\ P., Aleksandrovich N.\ L., Sunyaev R.\ A., Skinner G.\ K., 
1998, Astr.L. 24, 435 

\bibitem[Borozdin et al. 1999]{bor99} Borozdin K., Revnivtsev M., Trudolyubov 
S., Shrader C., Titarchuk L., 1999, ApJ, accepted (astro-ph/9812442)

\bibitem[Chakrabarti \& Titarchuk 1995]{ct95} Chakrabarti S.\ K.,
Titarchuk L.\ G., 1995, ApJ, 452, 226

\bibitem[Chen, Shrader \& Livio 1997]{chen97} Chen W., Shrader C.\ R., 
Livio M., 1997, ApJ, 491, 312

\bibitem[Cui et al. 1998]{cui98} Cui W., Ebisawa K., Dotani T., Kubota A., 
1998, ApJ, 493, L75

\bibitem[Cui et al. 1999]{cui99} Cui W., Zhang S.\ N., Chen W., Morgan E.\ H., 
1999, ApJ, in press (astro-ph/9812308)

\bibitem[Dove et al. 1998]{dove98} Dove J.\ B., Wilms J., Nowak M.\ A., 
Vaughan B.\ A., Begelman M.\ C., 1998, MNRAS, 298, 729 

\bibitem[Ebisawa, Mitsuda \& Inoue 1989]{Ebisawa89} Ebisawa K., Mitsuda K.,
Inoue H., 1989, PASJ, 41, 519

\bibitem[Ebisawa et al. 1994]{Ebisawa94} Ebisawa K., Ogawa M., Aoki T., 
Dotani T., Takizawa M., Tanaka Y., Yoshida K., Miyamoto S., Iga S., 
Hayashida K., Kitamoto S., Terada K., 1994, PASJ, 46, 375 
    
\bibitem[Fender \& Stappers 1998]{fender} Fender R.\ P., Stappers B.\ W., 
1998, IAU Circ. 6937
 
\bibitem[Fox \& Lewin 1998]{qpos} Fox D., Lewin W., 1998, IAU Circ. 6964

\bibitem[Hjellming, Rupen \& Mioduszewski 1998]{vla1} Hjellming R.\ M., 
Rupen M.\ P., Mioduszewski A.\ J., 1998, IAU Circ. 6934
 
\bibitem[Hjellming et al. 1998a]{vla2} Hjellming R.\ M., Rupen M.\ P., 
Chigo F., Waltman E.\ B., Mioduszewski A.\ J., 1998a, IAU Circ. 6937

\bibitem[Hjellming et al. 1998b]{h98} Hjellming R.\ M., Rupen M.\ P.,
Mioduszewski A.\ J., Smith D.\ A., Harmon B.\ A., Waltman E.\ B., 
Chigo F.\ D., Pooley G.\ G., 1998b, American Astron.Soc.Meeting \#193, 103.08

\bibitem[Kaneda et al. 1997]{gc_spectrum} Kaneda H., Makishima K.,
Yamauchi S., Koyama K., Matsuzaki K., Yamasaki N.\ Y., 1997, ApJ, 491, 638 

\bibitem[Kitamoto et al. 1992]{kit92} Kitamoto S., Tsunemi H., Miyamoto S., 
Hayashida K., 1992, ApJ, 394, 609

\bibitem[Makishima et al. 1986]{mak86} Makishima K., Maejima Y., Mitsuda K.,
Brandt H.\ V., Remillard R.\ A., Tuohy I.\ R., Hoshi R., Nakagawa M., 1986, 
ApJ, 308, 635

\bibitem[Markwardt, Swank \& Taam 1999]{m99} Markwardt C.\ B., Swank J.\ H.,
Taam R.\ E., 1999, ApJ, in press (astro-ph/9901050)
 
\bibitem[Mitsuda et al. 1984]{m84} Mitsuda K., Inoue H., Koyama K.,
Makishima K., Matsuoka M., Ogawara Y., Suzuki K., Tanaka Y., Shibazaki N.,
Hirano T., 1984, PASJ, 36, 741

\bibitem[Miyamoto et al. 1991]{Miyamoto91} Miyamoto S., Kimura K., 
Kitamoto S., Dotani T., Ebisawa K., 1991, ApJ, 383, 784

\bibitem[Miyamoto et al. 1994]{Miyamoto94} Miyamoto S., Kitamoto S., 
Iga S., Hayashida K., Terada K., 1994, ApJ, 435, 398

\bibitem[Molteni, Sponholz \& Chakrabarti 1996]{msc96} Molteni D., 
Sponholz H., Chakrabarti S.\ K., 1996, ApJ, 457, 805

\bibitem[Revnivtsev, Emelyanov \& Borozdin 1999]{Revnivtsev99}
Revnivtsev M.\ G., Emelyanov A.\ N., Borozdin K.\ N., 1999, Astr.L.,
25, 350 (astro-ph/9810156)  

\bibitem[Revnivtsev, Borozdin \& Emelyanov 1999]{rev1803} Revnivtsev M.\ G., 
Borozdin K.\ N., Emelyanov A.\ N., 1999, A\&AL, 334, 25

\bibitem[Rupen, Hjellming \& Mioduszewski 1998]{vla_jet} Rupen M.\ P., 
Hjellming R.\ M., Mioduszewski A.\ J. 1998, IAU Circ. 8938
 
\bibitem[Shakura \& Sunyaev 1973]{ss73} Shakura N.\ I., Sunyaev R.\ A., 
1973, A\&A, 24, 337

\bibitem[Smith, Levine \& Wood 1998]{asm} Smith D.\ A., Levine A., Wood A., 
1998, IAU Circ. 6932
 
\bibitem[Strohmayer \& Marshall 1998]{pca} Strohmayer T., Marshall F.\ E.,
1998, IAU Circ. 6934

\bibitem[Sunyaev \& Truemper 1979]{sun79} Sunyaev R.\ A., Truemper J., 1979, 
Nat, 279, 506

\bibitem[Sunyaev et al. 1994]{sun94} Sunyaev R.\ A., Borozdin K.\ N., 
Aleksandrovich N.\ L., Arefiev V.\ A., Kaniovsky A.\ S., Efremov V.\ V., 
Maisack M., Reppin C., Skinner G.\ K., 1994, Astr.L., 20, 777

\bibitem[Tanaka \& Lewin 1995]{TL95} Tanaka Y., Lewin W.\ H.\ G., 1995, 
in X-ray Binaries, ed. W.\ Lewin, J.\ van Paradijs, E.\ van der Heuvel 
(Cambridge: Cambridge Univ. Press), 126

\bibitem[Tanaka \& Shibazaki 1996]{Tanaka96} Tanaka Y., Shibazaki N., 1996, 
ARAA, 34, 607

\bibitem[Takizawa et al. 1997]{Takizawa97} Takizawa M., Dotani T.,
Mitsuda K., Matsuba E., Ogawa M., Aoki T., Asai K., Ebisawa K.,
Makishima K., Miyamoto S., Iga S., Vaughan B., Rutledge R.\ E.,
Lewin W.\ H.\ G. 1997, ApJ, 489, 272

\bibitem[Titarchuk, Lapidus \& Muslimov 1998]{tlm98} Titarchuk L., Lapidus
I., Muslimov A., 1998, ApJ, 499, 315

\bibitem[Trudolyubov et al. 1996]{Trudolyubov96}Trudolyubov S.\ P., Gilfanov
M.\ R., Churazov E.\ M., Borozdin K.\ N., Aleksandrovich N.\ L., 
Sunyaev R.\ A., Khavenson N.\ G., Novikov B.\ S., Vargas M., Goldwurm A., 
Paul J., Denis M., Borrel V., Bouchet L., Jourdain E., Roques J.-P., 
1996, Astr.L., 22, 664 

\bibitem[Trudolyubov et al. 1998]{Trud98} Trudolyubov S., Gilfanov
M., Churazov E., Sunyaev R., Khavenson N., Dyachkov A., Tserenin I., 
Sukhanov K., Laurent P., Ballet J., Goldoni P., Paul J., Roques J.-P.,  
Borrel V., Bouchet L., Jourdain E., 1998, A\&A, 334, 895

\bibitem[Trudolyubov et al. 1999]{Trudolyubov99} Trudolyubov S., Churazov E.,
Gilfanov M., 1999, Astr.L., in press (astro-ph/9811449)
 
\bibitem[Vikhlinin, Churazov $\&$ Gilfanov 1994]{Vihl94} Vikhlinin A., 
Gilfanov M., Churazov E., 1994, A\&A, 287, 73 

\bibitem[Yamauchi \& Koyama 1993]{gc_lines} Yamauchi S., Koyama K., 
1993, ApJ, 404, 620

\bibitem[Zhang et al. 1995]{Zhang95} Zhang W., Jahoda K., Swank J.\ H., 
Morgan E.\ H., Giles A.\ B., 1995, ApJ, 449, 930

\end{thebibliography}
\end{document}